\documentclass[iop]{emulateapj}
\usepackage{amsmath}
\usepackage{rotating}

\shorttitle{THE SHAPE OF X-RAY CAVITIES}
\shortauthors{GUO}

\begin{document}
\bibliographystyle{apj} 

\title {The Shape of X-ray Cavities in Galaxy Clusters: Probing Jet Properties and Viscosity}

\author{Fulai Guo\altaffilmark{1,2}}

\altaffiltext{1}{ETH Z\"{u}rich, Institute for Astronomy, Wolfgang-Pauli-Strasse 27, CH-8093, Z\"{u}rich, Switzerland; fulai.guo@phys.ethz.ch}

\altaffiltext{2}{Zwicky Prize Fellow}

\begin{abstract}

X-ray observations of galaxy clusters have detected numerous X-ray cavities, evolved from the interaction of AGN jets with the intracluster medium (ICM) and providing compelling evidence for the importance of jet-mode AGN feedback. Here we argue for the physical importance of the cavity shape, which we characterize with two geometric parameters: radial elongation $\tau$ and top wideness $b$. We study the cavity shape with 16 hydrodynamic jet simulations in two representative clusters, and find that the shapes of young cavities are mainly determined by various jet properties. Our simulations successfully reproduce two observed types of young cavities elongated along either the jet ($\tau>1$; type-II) or perpendicular ($\tau\leq1$; type-I) direction. Bottom-wide type-I cavities are produced by very light internally-subsonic jets, while top-wide type-II cavities are produced by heavier, internally-supersonic jets, which may also produce center-wide cavities with $\tau\sim 1$ if the jets are only slightly supersonic. Bottom-wide type-II cavities can be produced by very light jets with very long durations and cylindrical cavities are produced by very light internally-supersonic jets. While not appreciably affecting the shapes of young cavities, viscosity significantly affects the long-term cavity evolution, suppressing both interface instabilities and the formation of torus-like morphology. We encourage observers to study the shapes of young and old X-ray cavities separately, the former probing the properties of AGN jets and the latter potentially probing the ICM viscosity level.

\end{abstract}


\section{Introduction}
\label{section:intro}

It is widely accepted that mechanical feedback from active galactic nuclei (AGNs) plays a key role in the evolution of galaxy clusters, suppressing cooling flows and the associated fast growth of central galaxies (e.g., \citealt{mcnamara07}; \citealt{mcnamara12}). One of the most compelling evidences comes from mounting detections of surface brightness depressions in X-ray images of galaxy clusters, often referred as ``X-ray cavities" or ``AGN bubbles". Many cavities are associated with radio jets and spatially coincident with radio lobes (e.g., \citealt{boehringer93}; \citealt{fabian02}; \citealt{birzan04}; \citealt{croston11}), confirming the common note that they are evolved from the interaction of AGN jets with the intracluster medium (ICM). 

Taking the AGN jet origin for X-ray cavities, they are widely used to estimate the energetics of mechanical AGN feedback. While travelling through the ICM, the radio jets displace and do $pdV$ work on the surrounding hot gas, producing X-ray cavities. The total energy required to create a cavity is around
\begin{eqnarray}
E_{\rm cav}=\frac{\gamma}{\gamma -1}pV {\rm ,} \label{eq1}
   \end{eqnarray}
where $p$ is the pressure of the gas surrounding the cavity, $V$ is the cavity's volume, and $\gamma$ is the ratio of specific heats of the gas inside the cavity ($\gamma=4/3$ for relativistic gas). As these quantities are relatively well understood, this method is currently the most reliable to measure the non-radiative energy output from AGN jets.

In this paper, we argue for the first time that in addition to the cavity's volume ($V$ in Equation 1), its shape is also important. X-ray cavities are produced by the interaction of AGN jets with the surrounding gas, and subsequently rise in the ICM. Thus the cavity shape potentially encodes important information about the properties of AGN jets and the ICM. 

The ICM viscosity significantly affects the long-term evolution of X-ray cavities, but as we will demonstrate in this paper, it does not appreciably affect the shape of young cavities, which may be used to probe the properties of AGN jets. AGN feedback events may be triggered as supermassive black holes (SMBHs) accrete hot or cold gas from various origins \citep{best12}. The hot ICM is the dominant baryon component in galaxy clusters, and cold gas has also been observed in central regions of many cool core clusters (e.g., \citealt{peterson06}). Furthermore, the accretion of hot gas is directly affected by its heating and cooling balance, and if cooling dominates over heating intermittently (see \citealt{guo14} for more discussions), the accretion of cooling hot gas alone could be in either hot or cold mode, as seen in both steady state models (\citealt{quataert00}; \citealt{mathews12}) and hydrodynamic simulations \citep{guo14}. Thus due to potential variations in accretion modes, the properties of AGN jets may vary substantially in different systems, leaving signatures in the shapes of X-ray cavities. 

The observed cavities are the line-of-sight projected X-ray surface brightness depressions on the sky. During the last decade, a large number of X-ray cavities have been detected by {\it Chandra} and {\it XMM-Newton} X-ray space telescopes (see \citealt{mcnamara07} for a review). While some cavities are nearly circular, many cavities are not, as seen in the cavity samples of \citet{rafferty06} and \citet{hl12}. To evaluate the cavity volume used in Equation (\ref{eq1}), observers often approximate X-ray cavities as ellipses (ellipsoids in 3-dimensional space). Almost all non-circular cavities in \citet{rafferty06} and \citet{hl12} are elongated along either the jet direction (defined as the radial direction from the cluster center to the cavity center) or the perpendicular direction (defined as the direction perpendicular to the jet axis), suggesting that X-ray cavities are not subject to significant rotation during their evolution in the ICM. Otherwise, a large number of the cavities would be elongated along random directions with respect to the jet direction. 

 \begin{figure}
   \centering
\plotone{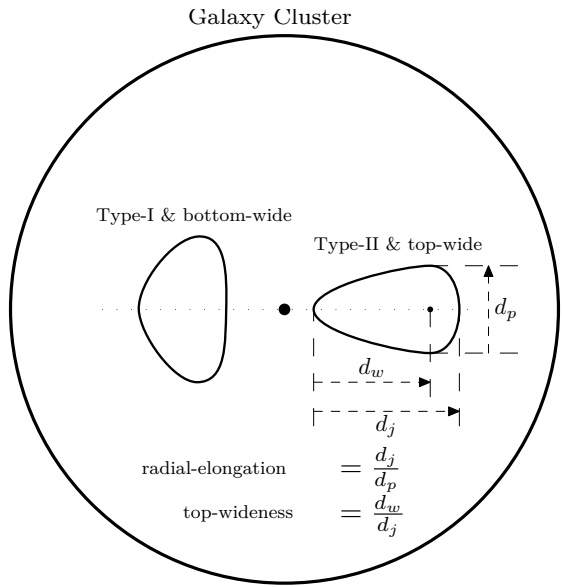} 
\caption{Sketch of two representative shapes of X-ray cavities in a galaxy cluster. We characterize the cavity shape with two parameters: radial elongation $\tau\equiv d_{\rm j}/d_{\rm p}$ and top wideness $b\equiv d_{\rm w}/d_{\rm j}$. The value of $\tau$ can be used to classify cavities into two types: type I elongated along the perpendicular direction ($\tau\leq 1$) and type II elongated along the jet direction ($\tau> 1$). The value of $b$ further separates each type of cavities into three subgroups: top-wide ($b>0.5$), center-wide ($b\sim 0.5$), and bottom-wide ($b<0.5$).}
 \label{plot1}
 \end{figure}

In this paper, we study the shape of X-ray cavities in hydrodynamic simulations, and particularly investigate the effects of jet properties and viscosity on the cavity shape. Our primary aim is to find some general trends connecting the cavity shape with some fundamental jet properties and the ICM viscosity, motivating future observational studies on this important topic. We describe our numerical setup in Section 2 and present the numerical results in Section 3. We conclude and discuss our results in Section \ref{section:discussion}. 

We characterize the cavity shape for the first time with two geometrical parameters $\tau$ and $b$. Radial elongation $\tau\equiv d_{\rm j}/d_{\rm p}$, defined as the ratio of the cavity axis along the jet direction ($d_{\rm j}$) to that along the perpendicular direction ($d_{\rm p}$), is a measure of the cavity elongation (the direction of the major axis) with respect to the jet axis. The value of $\tau$ can be used to classify cavities into two types: type I elongated along the perpendicular direction ($\tau\leq 1$) and type II elongated along the jet direction ($\tau> 1$). The second parameter, top wideness $b$, is a parameter to describe the deviation of a cavity's shape from a perfect ellipse by looking at the variation of its perpendicular size along the jet axis. The value of $b$ is defined as the relative location of the cavity's widest perpendicular size in its axis along the jet direction from the cavity bottom, $b\equiv d_{\rm w}/d_{\rm j}$, where $d_{\rm w}$ is the distance from the cavity bottom to the location of its widest perpendicular size in the jet axis. The value of $b$ further separates each type of cavities into three subgroups: top-wide ($b>0.5$), center-wide ($b\sim 0.5$), and bottom-wide ($b<0.5$). A simple sketch of two representative types of cavities is shown in Figure \ref{plot1}.
  
\section{Numerical Setup}
\label{section2}

 \begin{figure*}
   \centering
\plottwo{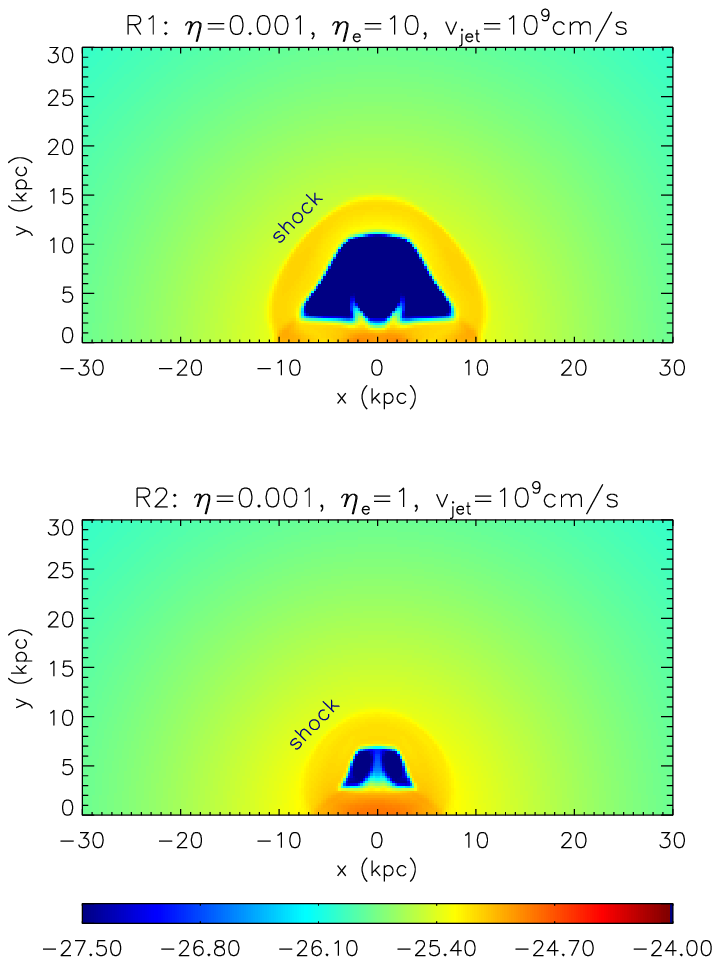} {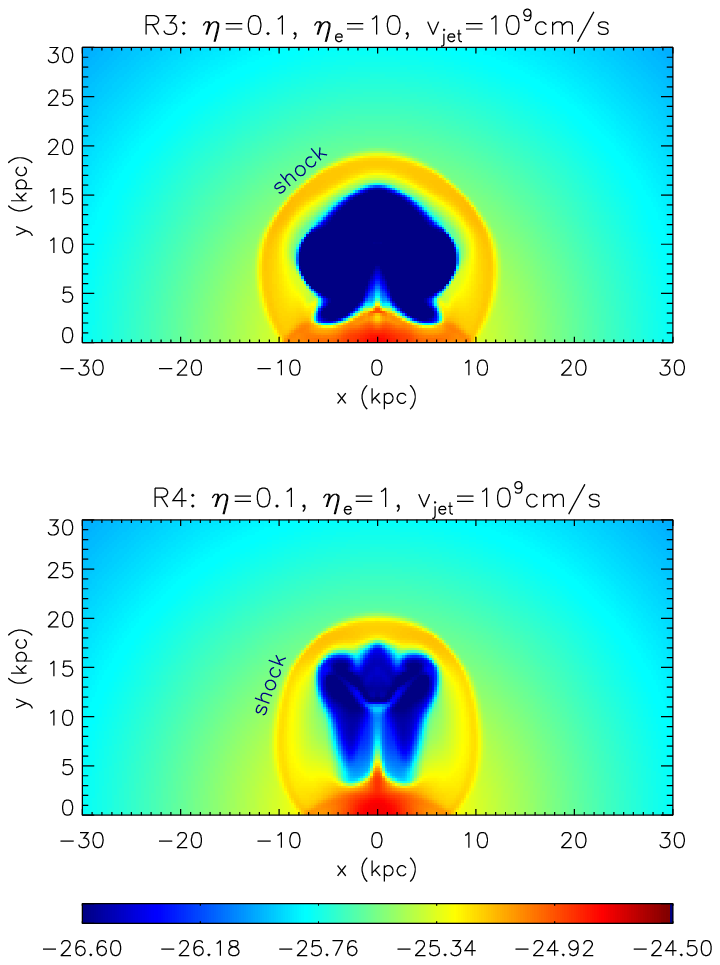} 
\caption{Central slices ($60 \times 30$ kpc) of gas density in logarithmic scale in a series of four simulations (with varying values of the initial jet density or energy density) at $t=t_{\rm jet}+2.5$ Myr. In each simulation, the jet is active during $0\leq t\leq t_{\rm jet}$. As seen in each panel, the jet produces a low-density cavity, surrounded by an outgoing bow shock further away from its surface. The color scale is saturated at the lowest density to better visualize the cavity.}
 \label{plot2}
 \end{figure*} 

 \begin{figure*}
   \centering
\plotone{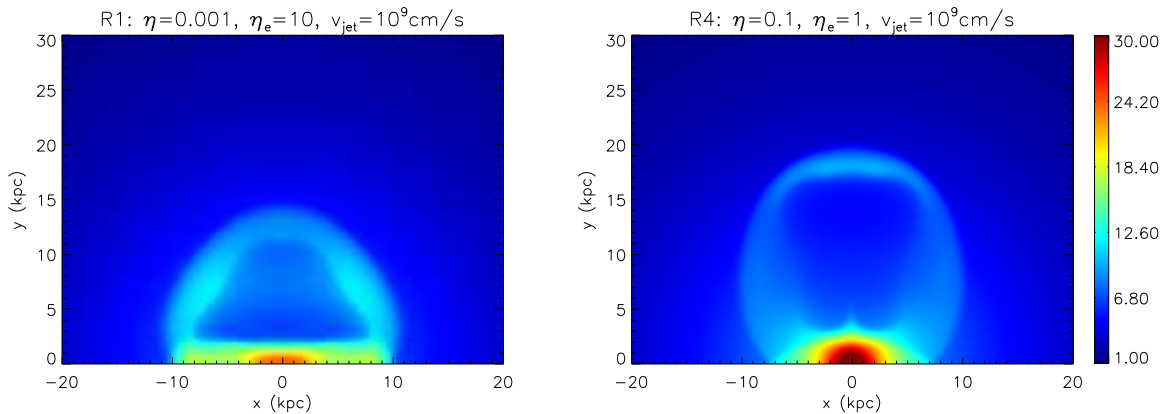} 
\caption{Synthetic X-ray surface brightness maps (line of sight projections of the cooling rate along a direction perpendicular to the jet axis in units of $10^{-4}$ erg cm$^{-2}$) of the central regions of the Virgo cluster in two typical jet simulations at $t=t_{\rm jet}+2.5$ Myr. The type-I cavity in run R1 (left panel) is produced by a jet with low density $\eta$ and high energy density $\eta_{\rm e}$, while the type-II cavity in run R4 (right panel) is produced by a jet with high density $\eta$ and low energy density $\eta_{\rm e}$.}
 \label{plot3}
 \end{figure*}

We consider the formation and evolution of X-ray cavities in the hot intracluster medium by a pair of bipolar jets released from the cluster center. Including viscosity, the basic hydrodynamic equations governing the gas evolution in the three-dimensional Eulerian space may be written as 
\begin{eqnarray}
\frac{d \rho}{d t} + \rho \nabla \cdot {\bf v} = 0,\label{hydro1}
\end{eqnarray}
\begin{eqnarray}
\rho \frac{d {\bf v}}{d t} = -\nabla P-\rho \nabla \Phi +\nabla \cdot {\bf \Pi},\label{hydro2}
\end{eqnarray}
\begin{eqnarray}
\frac{\partial e}{\partial t} +\nabla \cdot(e{\bf v})=-P\nabla \cdot {\bf v}+{\bf \Pi}:\nabla {\bf v}
   \rm{ ,}\label{hydro3}
   \end{eqnarray}
  \\ \nonumber
\noindent
where $d/dt \equiv \partial/\partial t+{\bf v} \cdot \nabla $ is the Lagrangian time derivative, {\bf $\Pi$} is the viscous stress tensor
\begin{eqnarray}
\Pi_{\rm ij}=\mu_{\rm visc}\left(\frac{\partial v_{\rm i}}{\partial x_{\rm j}}+\frac{\partial v_{\rm j}}{\partial x_{\rm i}}-\frac{2}{3}\delta_{\rm ij}\nabla \cdot {\bf v}\right)
 {\rm ,}
   \end{eqnarray}
$\mu_{\rm visc}$ is the dynamic viscosity coefficient, $\rho$ is the gas density, $P$ is the gas pressure, $e$ is the gas energy density, $v$ is the gas velocity, and $\Phi$ is the gravitational potential. We adopt the ideal gas law for the hot gas and the gas pressure and energy density are related via $P=(\gamma-1)e$, where $\gamma=5/3$. The molecular weight is assumed to be $\mu=0.61$. We ignore radiative cooling, which is unimportant during the short jet duration ($\lesssim 10$ Myr) in our simulations and may be offset by heating sources (e.g., AGN feedback or thermal conduction) on longer timescales. 

The simplest picture of the cavity formation contains a rotational symmetry with respect to the jet axis. We thus solve the above equations in $(r, z)$ cylindrical coordinates using a two-dimensional Eulerian code similar to ZEUS 2D \citep{stone92}. The code has also been successfully used in many previous studies, e.g., \citet{guo11}, \citet{guo12}, \citet{guo12b}. Our numerical implementation of shear viscosity has been described in detail in the Appendix of \citet{guo12b}. The computational grid consists of $400$ equally spaced zones in both coordinates out to $100$ kpc plus additional $100$ logarithmically-spaced zones out to $1$ Mpc. We have repeated several simulations with double resolution, and find that the shapes of young cavities converge quite well. The long-term evolution of cavities converges relatively well in viscous runs, but in non-viscous runs, old cavities are more disrupted in higher-resolution simulations, as discussed in more detail in \citet{bruggen01}. We adopt ``zero-gradient" boundary conditions at the outer boundary and reflective boundary conditions at the inner boundary. 

Our method is generally applicable to all reasonably relaxed clusters, but for concreteness, we adopt simulation parameters appropriate for the well-observed nearby system -- M87 and its halo within the Virgo cluster. The initial radial profiles of gas density and temperature are taken from the analytic fits to the observations by \citet{ghizzardi04}. The static gravitational field is determined to establish exact hydrostatic equilibrium for the initial gas pressure profile. The same initial cluster setup has been used in \citet{guo11}, where the readers are referred to for more details. In Section \ref{section:perseus}, we applied our method to a more massive cluster --- Perseus, and tested the robustness of our results.

The jet is introduced along the $z$-axis with the radius $r_{\rm jet}$ and an initial opening angle of $0$ degree. We initialize the jet at $z=z_{\rm jet}$ by adding gas fluxes of mass, energy, and momentum corresponding to a uniform jet with density $\rho_{\rm jet}$, energy density $e_{\rm jet}$ and velocity $v_{\rm jet}$. Real AGN jets on sub-kiloparsec scales ($z<1$ kpc) often move at relativistic velocities and are too narrow to be resolved by our numerical simulations. Thus we initialize jets in our simulations at $z_{\rm jet}=3$ kpc, and explore the dependence of the resulting cavity shape on $r_{\rm jet}$, $v_{\rm jet}$, and other jet parameters. Similar off-center jet injection methods have also been adopted in previous studies (e.g., \citealt{omma04}; \citealt{pm07}). Although real AGN jets may contain internal structures, our uniform jet setup allows us to study the connection of the cavity shape with simple jet parameters in great details.

The jet density and energy density can be normalized to the ambient ICM gas density $\rho_{\rm amb}$  and energy density $e_{\rm amb}$ at the jet base (here $\rho_{\rm amb}=1.43\times 10^{-25}$ g/cm$^3$ and  $e_{\rm amb}=5.55\times 10^{-10}$ erg/cm$^3$ at $z=z_{\rm jet}$ along the jet axis) by a density contrast $\eta\equiv \rho_{\rm jet}/\rho_{\rm amb}$ and energy density contrast $\eta_{\rm e}\equiv e_{\rm jet}/e_{\rm amb}$, respectively. An important jet parameter affecting its evolution is the internal Mach number at the jet base, which is defined as $M_{\rm int}=v_{\rm jet}/c_{\rm s,jet}=v_{\rm jet}/\sqrt{\gamma(\gamma-1)e_{\rm jet}/ \rho_{\rm jet}} $. The jet is active for a duration of $t_{\rm jet}$ and is turned off afterwards. To explore the parameter space, we presented in this paper a series of fourteen representative simulations, four of which include nonzero shear viscosity ($\mu_{\rm visc}$; see Section 3.2). The jet in each simulation injects both thermal and kinetic energies into the ICM. The kinetic power can be written as $P_{\rm ki}=\rho_{j} v_{\rm jet}^{3}\pi r_{\rm jet}^{2}/2$, and the thermal power is $P_{\rm th}=e_{ \rm j} v_{\rm jet}\pi r_{\rm jet}^{2}$. The jet parameters, energetics, and the shapes of resulting young cavities (values of $\tau$ and $b$) in all our simulations are listed in Table 1.

\begin{table*}
 \centering
 \begin{minipage}{160mm}
  \renewcommand{\thefootnote}{\thempfootnote} 
  \caption{List of Simulations}
    \vspace{0.1in}
  \begin{tabular}{@{}lccccccccccc}
  \hline & {$\eta$\footnote{$\eta=\rho_{\rm jet}/\rho_{\rm amb}$ is the initial jet density normalized by the ambient ICM gas density at the jet base.} }& {$\eta_{\rm e}$\footnote{$\eta_{\rm e}=e_{\rm jet}/e_{\rm amb}$ is the initial jet energy density normalized by the ambient ICM energy density at the jet base.} } &
         $v_{\rm jet}$& $t_{\rm jet}$&$r_{\rm jet}$&$\mu_{\rm visc}$& {$P_{\rm ki}$\footnote{$P_{\rm ki}=\rho_{j} v_{\rm jet}^{3}\pi r_{\rm jet}^{2}/2$ is the kinetic power of the jet.}} & {$P_{\rm th}$\footnote{$P_{\rm th}=e_{ \rm j} v_{\rm jet}\pi r_{\rm jet}^{2}$ is the thermal power of the jet.}}&$M_{\rm int}$&$\tau_{\rm young}$\footnote{$\tau_{\rm young}$ is radial elongation $\tau$ of the young cavity formed in each simulation at $t=t_{\rm jet}+2.5$ Myr.}&$b_{\rm young}$\footnote{$b_{\rm young}$ is tope wideness $b$ of the young cavity formed in each simulation at $t=t_{\rm jet}+2.5$ Myr. The shapes of young cavities in runs Rv1 and Rv2 are better described to be `cylindrical' than bottom-wide.}\\ Run&
         & &($10^{9}$cm/s)&(Myr)&(kpc)&($\text{  g cm}^{-1}\text{ s}^{-1}$)&(erg/s)&(erg/s)&\\ \hline 
         R1 &$0.001$&$10$& 1 &5&1&0& $2.14\times10^{42}$&$1.65\times10^{44}$&0.15&0.6&0.1\\ 
             R2 &$0.001$&$1$& 1 &5&1&0& $2.14\times10^{42}$&$1.65\times10^{43}$&0.48&0.5&0\\      
              R3 &$0.1$&$10$& 1 &5&1&0& $2.14\times10^{44}$&$1.65\times10^{44}$&1.5&0.9&0.5\\                  
         R4& $0.1$& $1$ & 1&5 &1&0&$2.14\times10^{44}$&$1.65\times10^{43}$&4.8&1.2&0.8\\
          Rv1  &$0.001$&$1$& 4.64 &5&1&0& $2.13\times10^{44}$&$7.67\times10^{43}$&2.2&0.9&cylindrical\\       
          Rv2  &$0.001$&$1$&10 &5&1&0& $2.14\times10^{45}$&$1.65\times10^{44}$&4.8&1.1&cylindrical\\       
          Rr1  &$0.001$&$10$& 1 &5&0.5&0& $5.34\times10^{41}$&$4.13\times10^{43}$&0.15&0.7&0\\       
          Rr2  &$0.001$&$10$&1 &5&2&0& $8.54\times10^{42}$&$6.61\times10^{44}$&0.15&0.5&0.2\\ 
            Rt1  &$0.001$&$10$& 1 &10&1&0& $2.14\times10^{42}$&$1.65\times10^{44}$&0.15&0.7&0.1\\      
          Rt2  &$0.001$&$10$&1&30&1&0&$2.14\times10^{42}$&$1.65\times10^{44}$&0.15&1.1&0.1\\            
         R1-visc1 &$0.001$&$10$& 1 &5&1&10& $2.14\times10^{42}$&$1.65\times10^{44}$&0.15&0.6&0.1\\ 
         R1-visc2 &$0.001$&$10$& 1 &5&1&100& $2.14\times10^{42}$&$1.65\times10^{44}$&0.15&0.6&0.1\\ 
         R4-visc1& $0.1$& $1$ & 1&5 &1&10&$2.14\times10^{44}$&$1.65\times10^{43}$&4.8&1.2&0.8\\
         R4-visc2& $0.1$& $1$ & 1&5 &1&100&$2.14\times10^{44}$&$1.65\times10^{43}$&4.8&1.2&0.8\\       
          \hline
\label{table1}
\end{tabular}
\end{minipage}
\end{table*}

 \begin{figure*}
   \centering
\plottwo{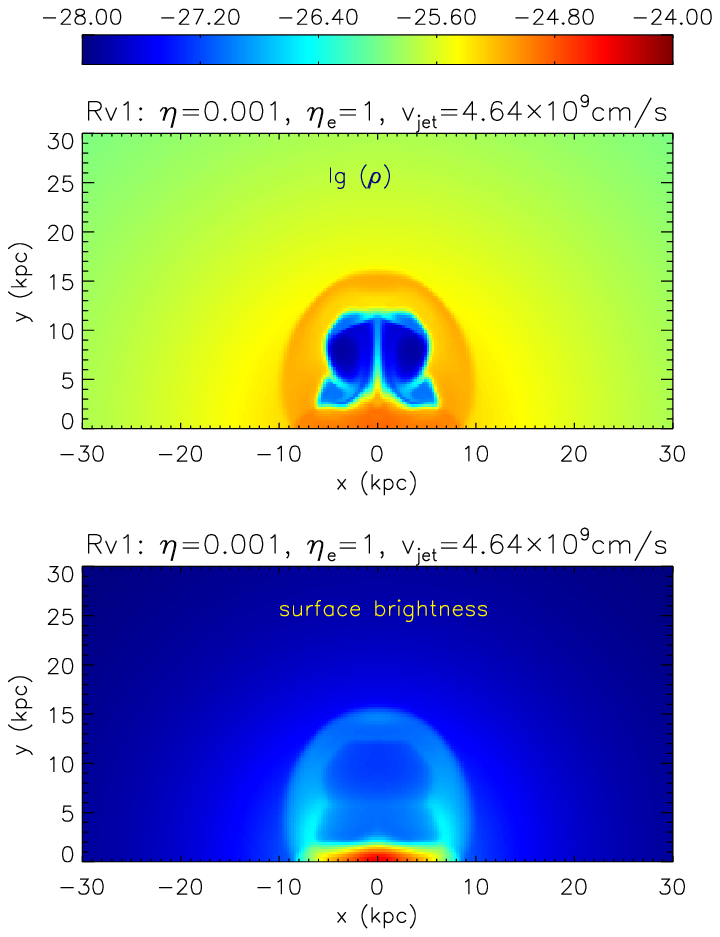}{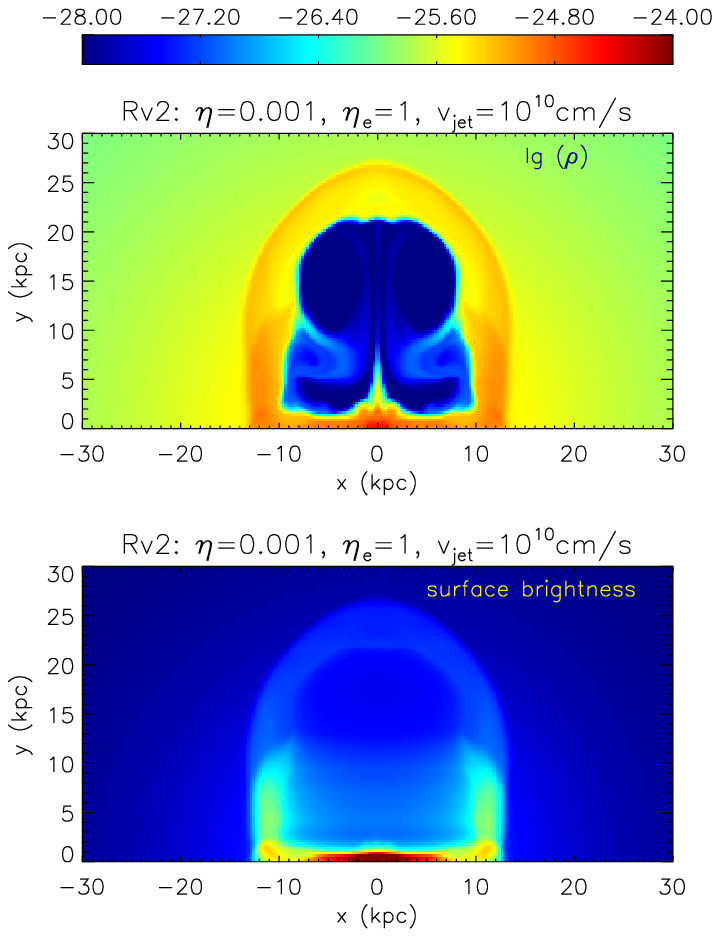} 
\caption{Impact of jet velocity on the cavity shape. Central slices of gas density (top panels) and synthetic X-ray surface brightness maps (bottom panels) are shown for runs Rv1 (left panels) and Rv2 (right panels) at $t=t_{\rm jet}+2.5$ Myr. The top color bar refers to gas density. The cavity becomes more radially elongated (i.e., the value of $\tau$ increases) as the jet velocity increases.}
 \label{plot4}
 \end{figure*} 

 \begin{figure*}
   \centering
\plottwo{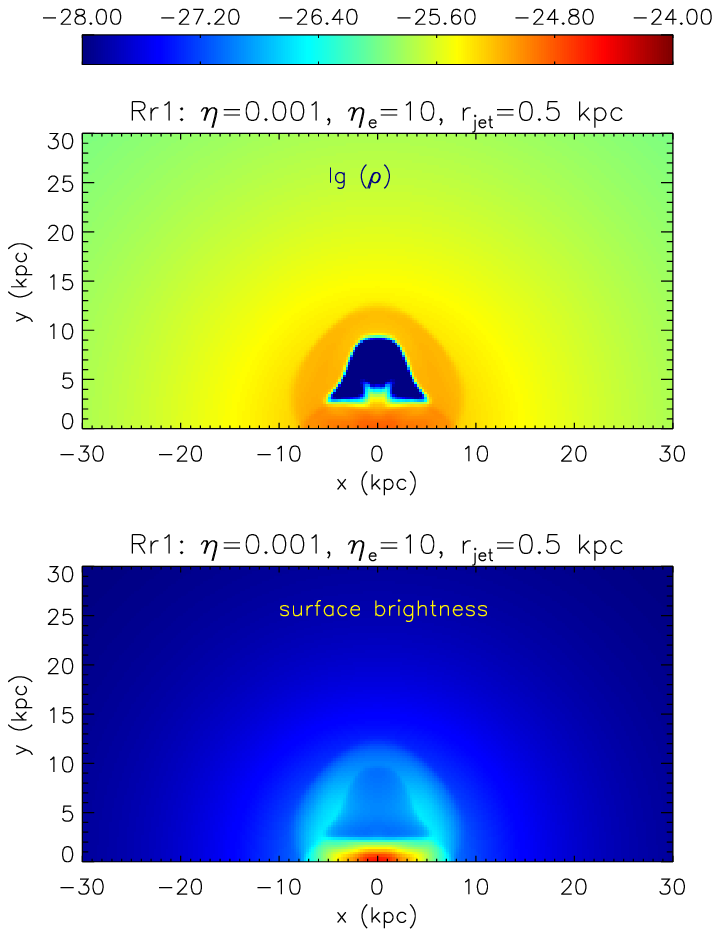}{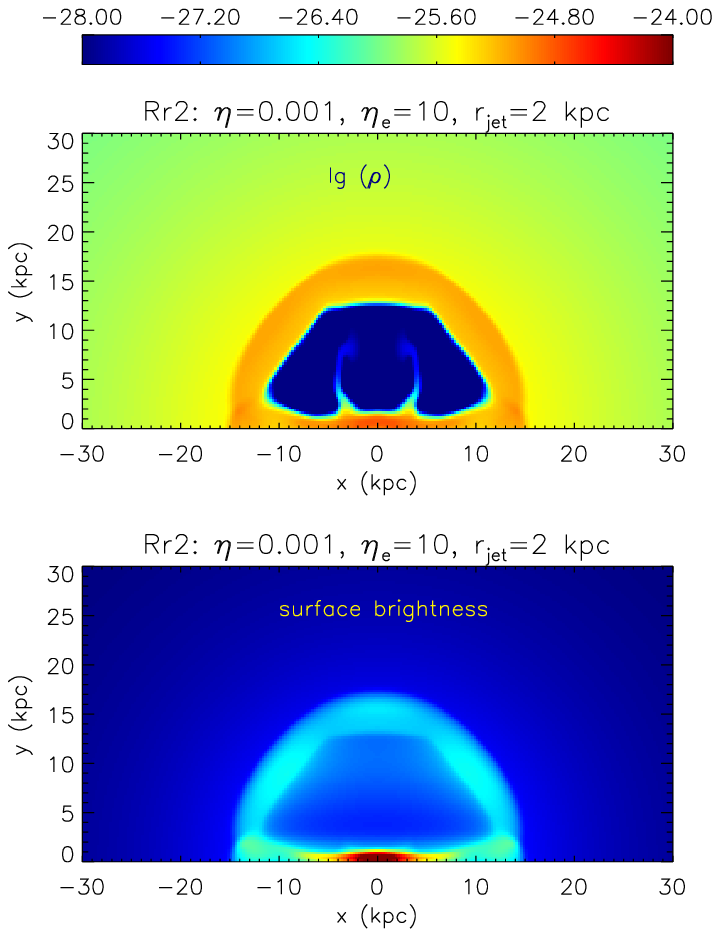} 
\caption{Impact of jet radius on the cavity shape. Central slices of gas density (top panels) and synthetic X-ray surface brightness maps (bottom panels) are shown for runs Rr1 and Rr2 at $t=t_{\rm jet}+2.5$ Myr. The cavity becomes more elongated along the perpendicular direction (i.e., the value of $\tau$ decreases) as the jet radius increases.}
 \label{plot5}
 \end{figure*} 

 \begin{figure*}
   \centering
\plottwo{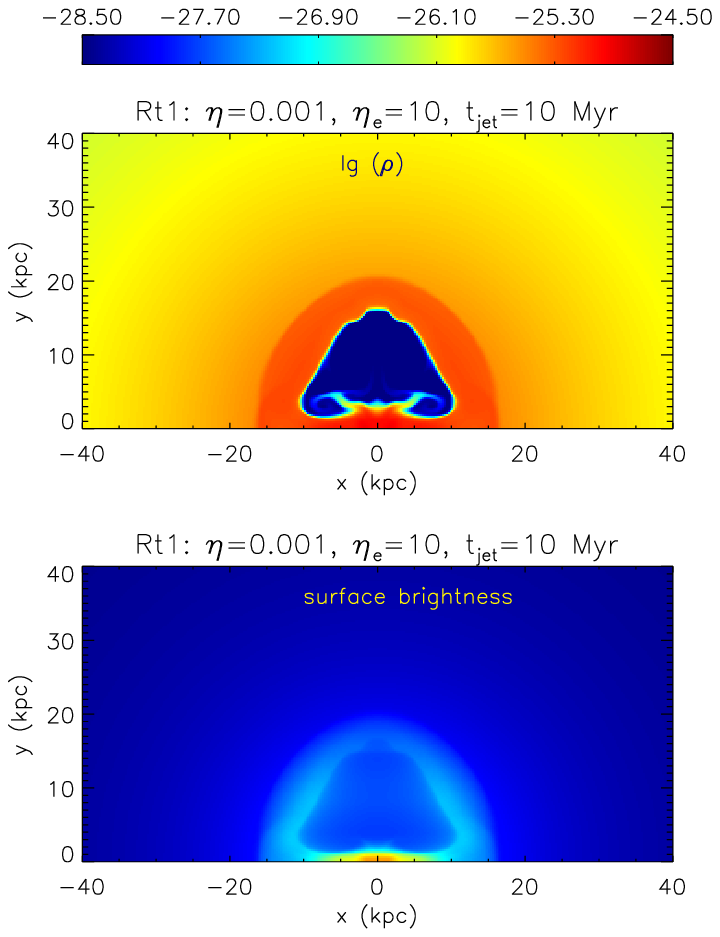}{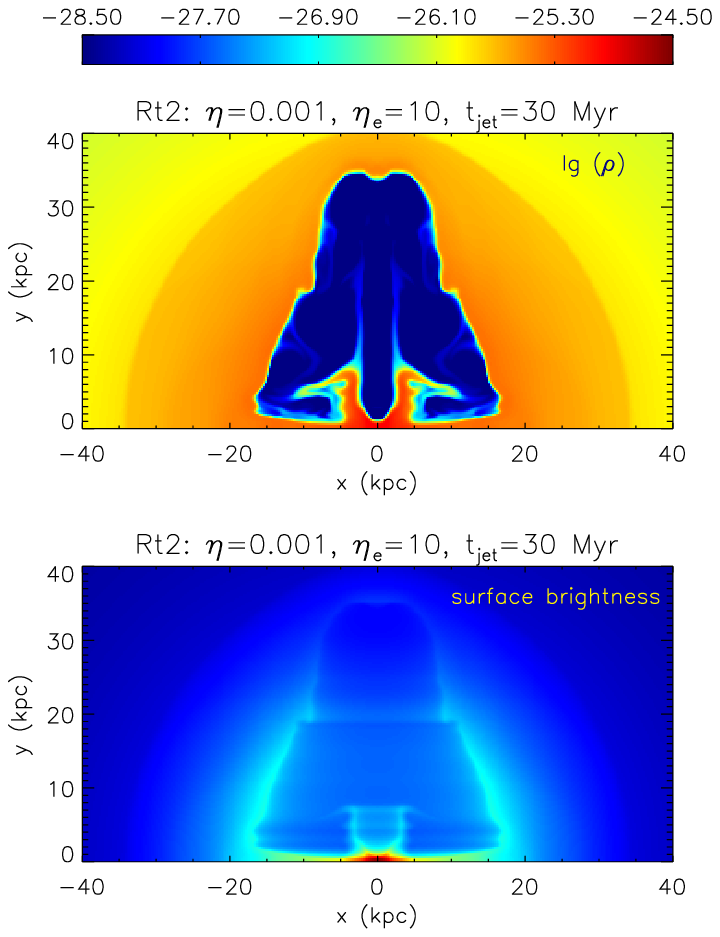} 
\caption{Impact of jet duration on the cavity shape. Central slices of gas density (top panels) and synthetic X-ray surface brightness maps (bottom panels) are shown for runs Rt1 and Rt2 at $t=t_{\rm jet}+2.5$ Myr. The cavity becomes more radially elongated (i.e., the value of $\tau$ increases) as the jet duration $t_{\rm jet}$ increases.}
 \label{plot6}
 \end{figure*} 
 
\section{The Cavity Shape in Hydrodynamic Simulations}
\label{section:simres}

The shape of X-ray cavities is determined by the properties of the jet and the ICM. In this paper we focus on the dependence of the cavity shape on the jet properties (jet parameters listed in Table 1) and the ICM viscosity. After the jet was turned off, the cavity rises away from the cluster center due to buoyancy and possibly the momentum injected by the original jet.  At this late stage, the cavity shape still changes with time. In Section 3.1, we focus on the shape of young cavities in our non-viscous simulations, which is directly related to the jet parameters. We then study the shape evolution of X-ray cavities in the ICM in Section 3.2, where the role of shear viscosity is also explored.

We note that the cavity shape discussed in our hydrodynamic simulations is ``intrinsic". For real observed cavities, the projection effect may make them appear more circular (i.e., with the value of $\tau$ closer to 1), as also discussed in \citet{birzan04}.

\subsection{The Shape of Young Cavities}
\label{section:young}

X-ray cavities are produced by AGN jets. We consider young cavities to be those that have not significantly risen away from their initial locations where they are created. In particular, here we study the shape of young X-ray cavities in our simulations at $t=t_{\rm jet}+2.5$ Myr, shortly after the jets are turned off at $t=t_{\rm jet}$. We do not study the early evolution of AGN jets and radio lobes, which have been explored in some previous studies (e.g., \citealt{krause05}; \citealt{gaibler09}).

We first investigate how the cavity shape depends on the jet density and energy density. Figure \ref{plot2} shows central slices of gas density in a series of four simulations at $t=t_{\rm jet}+2.5$ Myr with varying values of the initial jet density $\rho_{\rm jet}$ or energy density $e_{\rm jet}$. In each panel, clearly seen are a cavity (a 10$-$20-kpc-size region with relatively lower gas densities compared to the ambient value) and a surrounding bow shock (the outer discontinuity annotated in each panel) produced by the AGN jet event. Due to the high sound speed in the hot ICM, the bow shock is usually weak, with the Mach number in our fiducial run (run R1) dropping from $M\sim 1.6$ at $t=7.5$ Myr to $M\sim 1.2$ at $t=30$ Myr. 

The jet in run R1 is very light with density contrast $\eta=0.001$, and is over-pressured with respect to the ambient gas by a factor of $\eta_{\rm e}=10$. The R1 jet is internally subsonic with $M_{\rm int}=0.15$. The resulting cavity (the top left panel) is elongated more significantly along the perpendicular direction than the jet direction with the value of $\tau \sim0.6$. However, the cavity is not exactly an ellipse. If one looks at its perpendicular size $l_{\rm r}$ measured along the direction perpendicular to the jet axis, it is clear that  $l_{r}$ drops with the distance $r$ from the cluster center along the jet axis. The value of top wideness is $b\sim 0.1$, corresponding to a ``bottom-wide" type-I cavity in our classification as shown in Figure 1. This is an important feature, which could be used to disentangle different jet parameters for cavities with similar values of $\tau$ as further discussed below. The very light internally-subsonic jet in run R2 also produces a similar bottom-wide type-I cavity (the bottom left panel), which is much smaller than the R1 cavity due to the much less powerful jet in run R2.

In right panels, the jets in runs R3 and R4 are internally supersonic with $M_{\rm int}=1.5$ and $4.8$ respectively, and are more massive ($\eta=0.1$), resulting in cavities with higher values of radial elongation ($\tau \sim 0.9$ and $1.2$ respectively) than in runs R1 and R2. The R3 cavity is type-I and center-wide ($b\sim0.5$). Notably, the R4 cavity is elongated along the jet direction (a type-II cavity) and its perpendicular size roughly increases with the distance from the cluster center ($b\sim 0.8$; ``top-wide"). Thus Figure \ref{plot2} indicates that ``bottom-wide" type-I cavities can be produced by very light internally-subsonic jets, while ``top-wide" type-II cavities can be produced by heavier jets with internally supersonic velocities.

In Figure \ref{plot3}, we show the synthetic X-ray surface brightness maps of two typical jet simulations (runs R1 \& R4) at $t=t_{\rm jet}+2.5$ Myr. The X-ray surface brightness is computed as the line of sight projection of the cooling rate $n_{\rm e}n_{\rm i}\Lambda(T,Z)$ along a direction perpendicular to the jet axis. Here $n_{\rm e}$ is the electron number density, $n_{\rm i}$ is the ion number density, and $\Lambda(T,Z)$ is the \citet{sd93} cooling function at the gas metallicity $Z=0.4Z_{\sun}$. Type I and II cavities are clearly seen in the left (run R1) and right (run R4) panels, respectively. The internally-subsonic jet in run R1 is energetically dominated by the thermal power, and the internally-supersonic jet in run R4 is energetically dominated by the kinetic power (see Table 1). Type-I cavities can also be produced by very light jets energetically dominated by cosmic rays, as seen in \citet{guo11}. 

We then study how the cavity shape depends on the jet velocity in runs Rv1 and Rv2, as shown in Figure \ref{plot4}. The jet parameters in these two runs are identical to those in run R2 except for higher values of the jet velocity. Comparing the bottom left panel of Figure \ref{plot2} and Figure \ref{plot4}, it is clear that the cavity becomes more radially elongated (i.e., the value of $\tau$ increases) as the jet velocity increases. Although type II cavities can be created by very light jets with high jet velocities (e.g., run Rv2 with $\eta=0.001$, $v_{\rm jet}=10^{10}$ cm/s, and $M_{\rm int}=4.8$), these cavities are not ``top wide" as those created by heavier jets (e.g., in run R4 with $\eta=0.1$). In fact, the cavities produced by very light internally-supersonic jets in runs Rv1 and Rv2 are close to ``cylindrical", as also found in \citet{krause05} and a relativistic simulation in \citet{pm07}. In homogenous environments, very light internally-supersonic jets even produce bottom-wide cavities, as shown in \citet{gaibler09}.

The cavity shape also depends on the jet radius $r_{\rm jet}$ at the jet base, which is explored in runs Rr1 and Rr2 shown in Figure \ref{plot5}. The jet parameters in these two runs are identical to those in our fiducial run R1 except that $r_{\rm jet}$ is changed to $0.5$ and $2$ kpc in runs Rr1 and Rr2 respectively. The very light jets ($\eta=0.001$) in these two runs also produce ``bottom-wide" cavities as in run R1, and the cavity becomes more elongated along the perpendicular direction (i.e., the value of $\tau$ decreases) as the jet radius increases.
 
The dependence of the cavity shape on the jet duration $t_{\rm jet}$ is investigated in runs Rt1 and Rt2, which have the same jet parameters as in run R1 except that $t_{\rm jet}$ is increased to $10$ and $30$ Myr in runs Rt1 and Rt2 respectively. As shown in Figure \ref{plot6}, the cavity shape in run Rt1 is similar to that in run R1, but the cavity in run Rt2 is elongated along the jet direction with a much larger radial elongation $\tau$ than in run R1. Thus the very light jets in these two runs also produce ``bottom-wide" cavities as in run R1, but the cavity becomes more radially elongated as the jet duration increases.

In this section, we presented a suite of 10 simulations, showing that the shape of young X-ray cavities is significantly affected by various jet parameters, including $\rho_{\rm jet}$, $e_{\rm jet}$, $v_{\rm jet}$, $r_{\rm jet}$ and $t_{\rm jet}$. Our simulations explicitly demonstrated that it is feasible to produce both type-I and -II cavities by AGN jets and found some general trends. Very light jets (e.g., $\eta=0.001$ in run R1) produce very strong backflows, which transport jet materials away from jet heads. The back-flowed jet materials tend to accumulate near the bottom of cavities and expand laterally. If jets are very-light and internally subsonic, the forward motions of jet heads are relatively slow and the cavities' lateral expansion is often significant enough to produce bottom-wide cavities elongated along the perpendicular direction (type I). If these jets are active for a very long duration, they tend to create bottom-wide cavities elongated along the jet direction (type II). On the other hand, if the velocity of a very light jet is very large (internally supersonic), the jet head's motion tends to be comparable to or even more important than the cavity's lateral expansion. Therefore, the resulting cavity may also be elongated along the jet direction, but it  appears more cylindrical than bottom-wide. For heavier jets (e.g., the light jet with $\eta=0.1$ in run R4), the backflows are less strong, and the jet heads moves relatively faster, resulting in cavities with larger values of $\tau$ and $b$ (e.g., center-wide or top-wide). 

In summary, bottom-wide type-I cavities are produced by very light internally-subsonic jets, while top-wide type-II cavities are produced by heavier jets with internally supersonic velocities, which may also produce center-wide cavities with $\tau\sim 1$ if the jets are slightly supersonic ($M_{\rm int}\sim 1$-$2$). In our simulations, we adopt very light jets with $\eta=0.001$ and heavier jets with $\eta=0.1$. A general threshold $\eta_{\rm c}=0.01$ may be used to distinguish between very light ($\eta<0.01$) and heavier ($\eta>0.01$) jets. Bottom-wide type-II cavities can be produced by very light jets with very long durations ($t_{\rm jet}$). Cylindrical cavities (both type I and II) are produced by very light internally-supersonic jets, and their radial elongations increase (from type-I to type-II cavities) with the jet's internal Mach number $M_{\rm int}$.

 \begin{figure*}
   \centering
\plotone{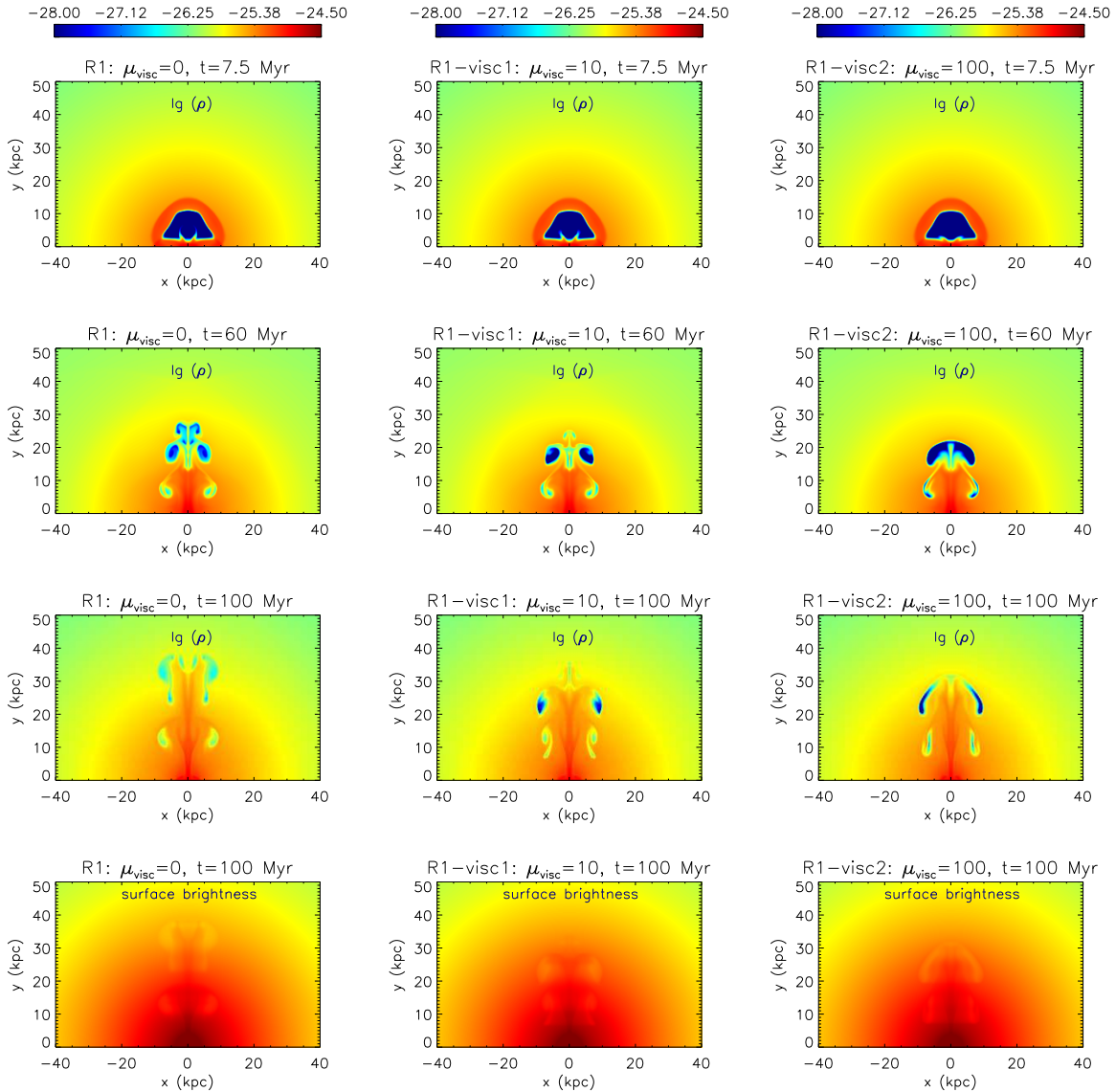} 
\caption{Temporal evolution of type-I cavities in three simulations with different viscosity coefficients ($\mu_{\rm visc}=0, 10, 100$ g cm$^{-1}$ s$^{-1}$ as indicated at the top of each panel). In each column, the upper three panels show central density slices at three times $t=7.5$, $60$, and $100$ Myr in each specific simulation, and the bottom panel shows the corresponding synthetic X-ray surface brightness map at $t=100$ Myr. The top color bar refers to gas density. Although not appreciably affecting the shape of young cavities, viscosity does affect the late evolution of X-ray cavities.}
 \label{plot7}
 \end{figure*} 

 \begin{figure*}
   \centering
\plotone{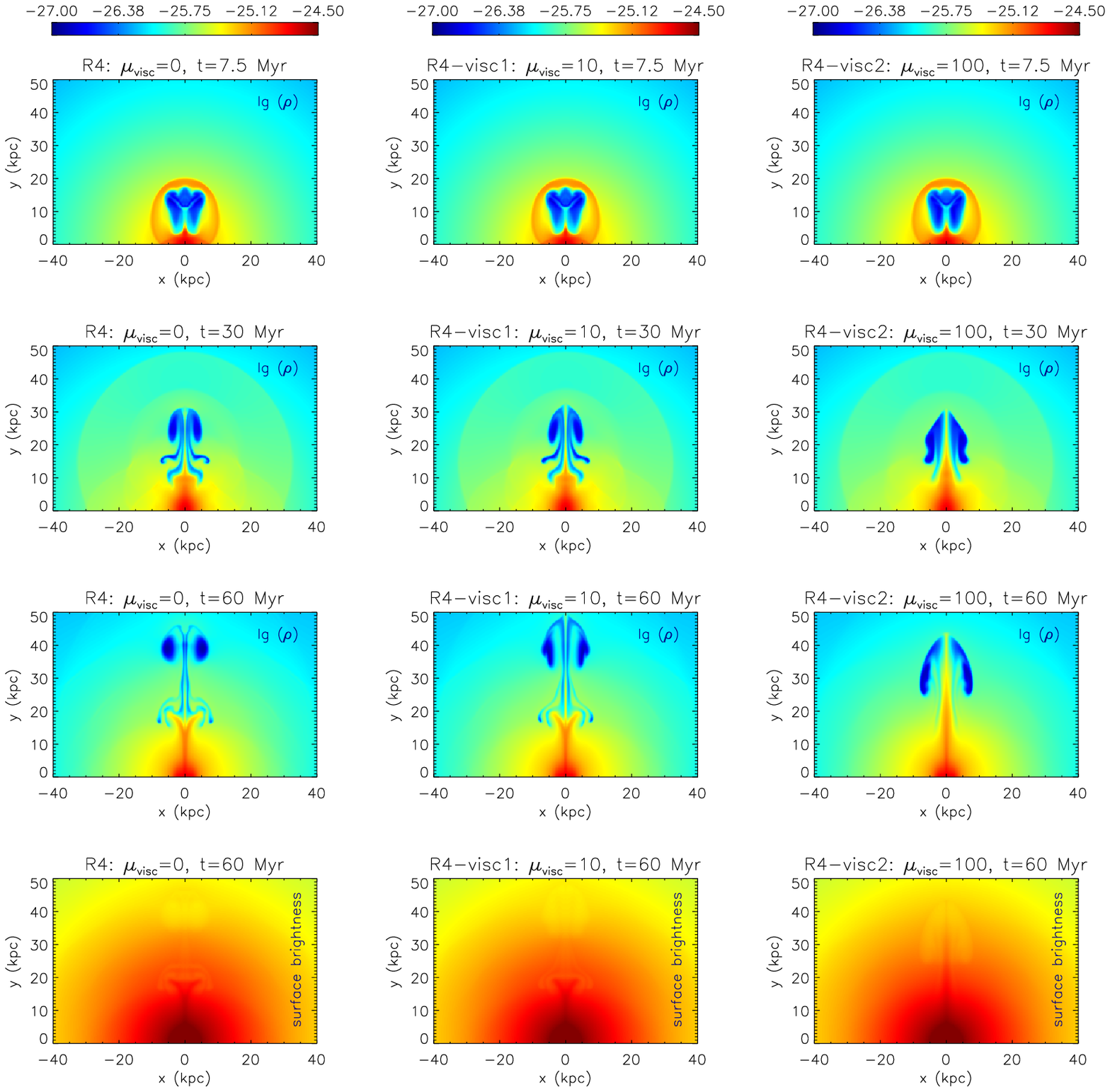} 
\caption{Temporal evolution of type-II cavities in three simulations with different viscosity coefficients. In each column, the upper three panels show central density slices at three times $t=7.5$, $30$, and $60$ Myr in each specific simulation, and the bottom panel shows the corresponding synthetic X-ray surface brightness map at $t=60$ Myr. Although not appreciably affecting the shape of young cavities (e.g. top panels), viscosity does affect the late evolution of X-ray cavities.}
 \label{plot8}
 \end{figure*}

\subsection{The Shape Evolution of Cavities in the ICM}
\label{section:young}

We study the late evolution of X-ray cavities and the role of shear viscosity in this subsection. After the jet is turned off, the cavity moves away from the cluster center (i.e., ``rises in the ICM") due to the combined effects of buoyancy and the original momentum injected by the jet. During this process, the cavity shape, and in particular its radial elongation $\tau$ and top wideness $b$, may evolve with time. However, it is not easy to accurately predict the shape evolution of cavities in this process, as they are subject to the disruption by Rayleigh-Taylor (RT) and Kelvin-Helmholtz (KH) instabilities developed at the cavity surface. This can be seen in the left panels of Figures \ref{plot7} and \ref{plot8} for runs R1 and R4 respectively, as also studied in \citet{reynolds05} for initially static cavities.

To keep the cavity intact (such as the outer cavities in the Perseus cluster observed by \citealt{fabian00}; also see \citealt{chon12} for another potential evidence in Cygnus A), additional physics, e.g., shear viscosity (\citealt{kaiser05}; \citealt{reynolds05}; \citealt{guo12b}) or magnetic tension (\citealt{kaiser05}; \citealt{jones05}; \citealt{ruszkowski07}), was proposed to suppress the instabilities. Here we briefly investigate the role of shear viscosity on the cavity evolution. Our simulations are the first set of simulations that study the evolution in the viscous ICM  of X-ray cavities {\it directly created by AGN jets}, while previous studies (e.g., \citealt{reynolds05}) follow the evolution of an initially-static cavity which neglects the backflows and the internal circulating motions produced during the interaction of AGN jets with the ICM.

Following \citet{reynolds05}, we introduce a constant dynamic viscosity coefficient $\mu_{\rm visc}$ in our four viscous runs: R1-visc1, R1-visc2, R4-visc1, R4-visc2 (see Table 1 for model parameters). The jet parameters in the first two runs are the same as those in our fiducial run R1 except that $\mu_{\rm visc}$ is increased from zero in run R1 to $10$ g cm$^{-1}$ s$^{-1}$ in run R1-visc1 and $100$ g cm$^{-1}$ s$^{-1}$ in R1-visc2. Similarly, runs R4-visc1 and R4-visc2 are very similar to run R4 except for the different values of $\mu_{\rm visc}$. As a comparison, the theoretical Braginskii viscosity in a  fully ionized, unmagnetized, thermal plasma is $\mu_{\rm visc}\sim 156(\text{ln}\Lambda/37)^{-1}(T/2\text{ keV})^{2.5}$ g cm$^{-1}$ s$^{-1}$ \citep{braginskii58}.

Figure \ref{plot7} shows the time evolution of type-I cavities in runs R1, R1-visc1, and R1-visc2. It is clear that viscosity does not appreciably affect the shape of young cavities (e.g., top panels at $t=7.5$ Myr). However, viscosity does affect the late evolution of X-ray cavities and suppresses the development of KH and RT instabilities, consistent with previous studies (e.g. \citealt{reynolds05}). The cavity evolution in run R1-visc2 (the right panels) suggests that type-I cavities becomes more elongated along the perpendicular direction (i.e., $\tau$ decreases) as they rise in the ICM.

Figure \ref{plot8} shows the time evolution of type-II cavities in runs R4, R4-visc1, and R4-visc2. Similar to type-I cavities, viscosity does not appreciably affect the shape of young type-II cavities, but does affect their long-term evolution by suppressing interface instabilities. Another important feature is that the cavities in these runs (Figure \ref{plot8}) rise faster than those in runs  R1, R1-visc1, and R1-visc2 (Figure \ref{plot7}), suggesting that higher momentum fluxes injected by more massive jets in the R4-series runs contribute significantly to the late cavity rise in the ICM. In other words, bottom-wide type-I cavities rise in the ICM mainly due to buoyancy, while top-wide type-II cavities may rise in the ICM due to the combined effects of buoyancy and the original momenta injected by AGN jets.

Figures \ref{plot7} and \ref{plot8} show that shear viscosity significantly changes the late evolution of X-ray cavities in runs R1-visc2 and R4-visc2, effectively suppressing RT and KH instabilities. Cavities in our non-viscous runs evolve into torus-like structures (as also seen in \citealt{pavlovski08}), while viscosity tends to suppress the formation of torus-like morphology, making cavities more coherent. The viscosity coefficients in these two runs are $\mu_{\rm visc}=100$ g cm$^{-1}$ s$^{-1}$, a significant fraction of the Braginskii viscosity ($\sim 156$ g cm$^{-1}$ s$^{-1}$ at $T=2$ keV). A level of $\mu_{\rm visc}=10$ g cm$^{-1}$ s$^{-1}$ in runs R1-visc1 and R4-visc1 produces a much smaller effect. 

Our 2D simulations assumed axisymmetry, and thus could not explore the roles of non-axisymmetric features potentially present in both AGN jets and the ambient ICM. As studied in \citet{krause05} in detail, non-axisymmetric motions may affect turbulent mixing within cavities and the development of interface instabilities, but we do not expect that they significantly affect the overall shape of young X-ray cavities and their long-term evolution in a viscous ICM. Future 3D viscous simulations including a small level of non-axisymmetry in the initial ICM setup (e.g., as adopted in \citealt{krause05}) are required to explore this interesting topic.

 \begin{figure*}
   \centering
\plottwo{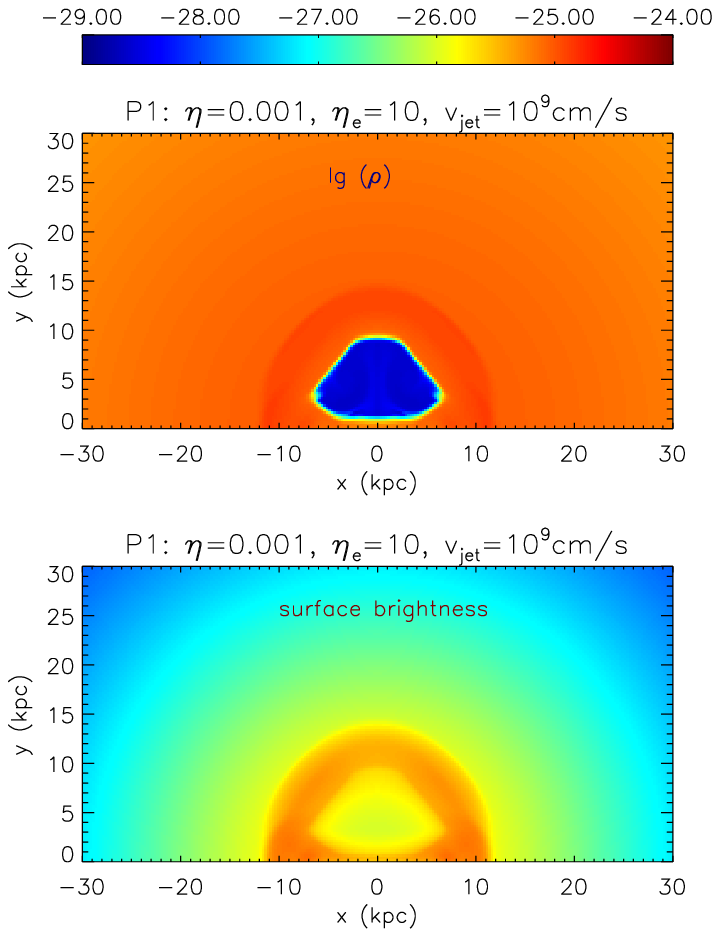}{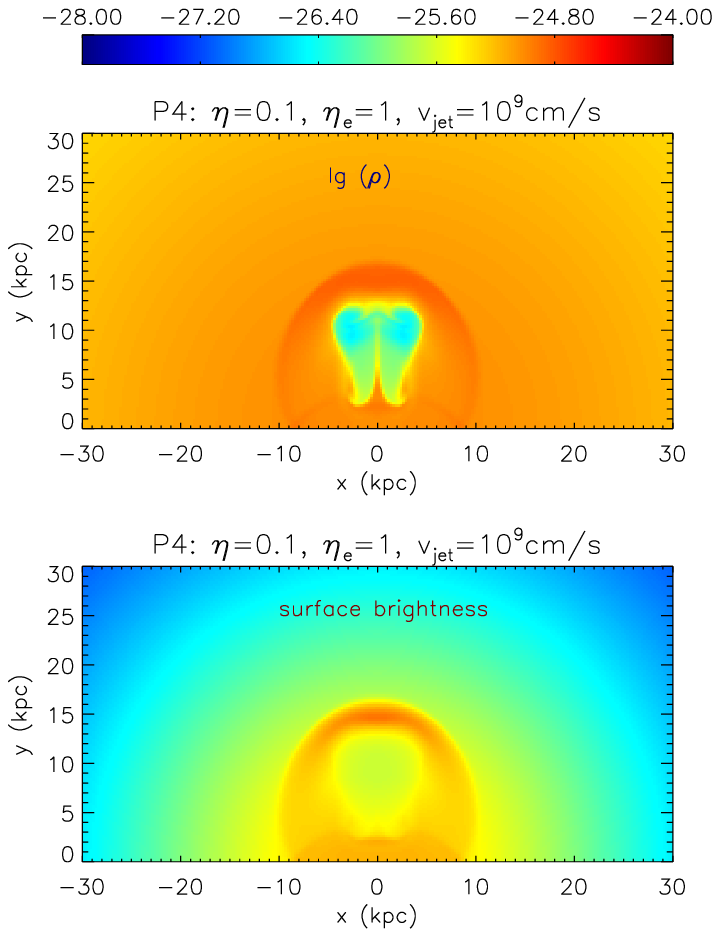} 
\caption{Formation of type I (run P1; left panels) and type II (run P4; right panels) X-ray cavities in the Perseus cluster. Central slices of gas density (top panels) and synthetic X-ray surface brightness maps (bottom panels) are shown for each run at $t=t_{\rm jet}+2.5$ Myr. The top color bar refers to gas density. }
 \label{plot9}
 \end{figure*} 
 
\subsection{Formation of Type I and II Cavities in the Perseus Cluster}
\label{section:perseus}

In this subsection, we explore the robustness of our results by studying the formation of type I and II X-ray cavities in another system --- the Perseus cluster. Perseus is much more massive than our fiducial system Virgo, and the ICM temperature is about 2 -- 3 times higher. The gas pressure in Perseus is also higher than that in Virgo (by a factor of $\sim3$ at $r=7$ kpc and $\sim 6$ at $r=32$ kpc according to observational profiles in \citealt{ghizzardi04} and \citealt{churazov04}).

Here we present two representative Perseus simulations, runs P1 and P4, producing type I and II cavities respectively. We use the same numerical method as described in Section \ref{section2}. The initial gas temperature and density profiles in Perseus are adopted from observational profiles given in \citet{churazov04}. The static gravitational potential well is also chosen to establish hydrostatic equilibrium at $t=0$. The jet parameters ($\eta$, $\eta_{\rm e}$, $v_{\rm jet}$, $t_{\rm jet}$, $r_{\rm jet}$) in runs P1 and P4 are the same as those in runs R1 and R4 (see Table 1), respectively.

Figure \ref{plot9} shows the cavities produced in runs P1 and P4 at $t=t_{\rm jet}+2.5$ Myr. The very light internally-subsonic jet in run P1 ($\eta=10^{-3}$, $M_{\rm int}=0.11$) produces a bottom-wide type-I cavity in Perseus (similar to run R1 for Virgo), while the heavier, internally-supersonic jet in run P4 ($\eta=0.1$, $M_{\rm int}=3.5$) produces a top-wide type-II cavity (similar to run R4). This confirms our main results presented in Section \ref{section:young}, suggesting that the connection between the shapes of young cavities and jet properties may be quite general. It should also be noted that the sizes of cavities in Perseus are slightly smaller than the sizes of corresponding Virgo cavities produced by the same jet properties, due to higher ICM gas pressures in Perseus.

X-ray observations indicate that the scaled angular-averaged pressure profiles in galaxy clusters are quite universal \citep{arnaud10}. We thus expect that the main results in the paper should hold quite generally. The shapes of young cavities, particularly radial elongation and top wideness, may be useful probes of the properties of AGN jets producing those cavities. However, we also caution that local variations in the surrounding gas pressure may lead to asymmetries and irregularities in the bubble shape. Angular variations in the ICM distribution and the differences in the jet entrainment history may cause that a pair of opposite jets produce two cavities with different shapes. These effects should be considered while inferring jet properties from the shapes of observed young X-ray cavities.

\subsection{Comparison with Previous Studies}
\label{section:comparison}

The propagation of AGN jets and the formation of radio lobes and X-ray cavities in the ICM have been previously studied with hydrodynamic simulations by many authors in the literature (e.g., \citealt{clarke97}; \citealt{reynolds01}; \citealt{reynolds02}; \citealt{basson03}; \citealt{zanni03}; \citealt{ruszkowski04}; \citealt{krause05}; \citealt{bruggen07}; \citealt{pavlovski08}; \citealt{guo10a}; \citealt{gaspari11}; \citealt{hardcastle13}), sometimes including additional physics, e.g., magnetic fields (\citealt{jones05}; \citealt{ruszkowski07}; \citealt{gaibler09}), cosmic rays \citep{guo11}, viscosity (\citealt{reynolds05}; \citealt{dong09}), etc.  Many previous studies focus on the thermodynamic impact of mechanical AGN feedback on the ICM (e.g., \citealt{bruggen07}; \citealt{guo10a}; \citealt{guo10b}; \citealt{gaspari14}; \citealt{li14}) and how to suppress the cavity disruption by interface instabilities (e.g., \citealt{kaiser05}; \citealt{reynolds05}; \citealt{jones05}; \citealt{ruszkowski07}). In the current paper, we present a very detailed and extensive study on the cavity shape produced by AGN jets with many different jet parameters, including very light internally-subsonic jets largely ignored in previous studies.  In particular, we propose two very important parameters to characterize the shape of X-ray cavities: radial elongation $\tau$ and top wideness $b$.

Our results are consistent with \citet{guo11}, both demonstrating that massive (high-density) jets tend to form cavities elongated along the jet direction and ``fat" type-I cavities are formed by very light jets. The major difference is that our jets producing type-I cavities are energetically dominated by thermal energy density, while the jets in \citet{guo11} are dominated by the cosmic ray energy density. \citet{sternberg07} showed that fat cavities can also be produced by massive jets with very large half-opening angles ($>50^{\circ}$), although radio jets are usually observed to be well collimated (e.g., \citealt{kovalev07}).

\citet{reynolds05} studied the long-term evolution of initially-static cavities in the ICM and the role of shear viscosity. Our results in Section 3.2 are generally consistent with \citet{reynolds05} on the role of viscosity in suppressing KH and RT instabilities. However, starting with initially-static cavities, \citet{reynolds05} could not probe the role of the jet momentum, which likely plays an important role in the rise of type II cavities in the ICM as shown in Figure \ref{plot8}. Our simulations directly probe the formation process of X-ray cavities by AGN jets, which produce jet backflows and circulating motions in cavities (e.g., \citealt{norman82}; \citealt{guo12b}). For type-I cavities shown in Figure \ref{plot7}, some backflows reaching the cluster central regions detach from the cavity at late times (e.g., at $t=60, 100$ Myr), even in run R1-visc2 with quite strong viscosity $\mu_{\rm visc}=100$ g cm$^{-1}$ s$^{-1}$. This feature is not seen in simulations of initially-static cavities \citep{reynolds05}, and it is unclear if it represents reality. The detached cavity material leads to a small X-ray-deficient area in X-ray surface brightness maps (bottom panels), and may not be easily seen in real X-ray images. If the non-existence of this feature is observationally confirmed, further investigations are necessary to explain it.

\section{Summary and Discussion}
\label{section:discussion}

{\it Chandra} and {\it XMM-Newton} observations of galaxy clusters have detected a large number of X-ray cavities in galaxy groups and clusters, providing one of the most compelling evidences for the importance of jet-mode AGN feedback (e.g.,  \citealt{mcnamara07}). Through the cavity's volume and the surrounding gas pressure, X-ray cavities have been successfully used to estimate the energetics of mechanical AGN feedback (e.g., \citealt{birzan04}; \citealt{rafferty06}; \citealt{osullivan11}). 

Here in this paper we argue for the physical importance of the shapes of X-ray cavities, whose importance has been largely ignored in previous studies. X-ray cavities are created by the interaction of AGN jets with the surrounding ICM gas, and thus the cavity shape potentially contains important information about the properties of AGN jets and the ICM. In particular, the cavity elongation with respect to the jet direction may be of great interest. X-ray observations indicate that most cavities are elongated along either the perpendicular direction (including nearly circular cavities) or the jet direction (e.g., \citealt{rafferty06}; \citealt{hl12}), suggesting that X-ray cavities are usually not subject to substantial rotation during their evolution in the ICM. This further implies that typical turbulent motions in galaxy clusters do not effectively rotate the cavities. However, it should be noted that coherent gas motions (e.g., gas sloshing) may bend (or rotate) radio lobes and X-ray cavities (e.g., \citealt{pm13}; \citealt{venturi13}), as numerically simulated in \citet{mendygral12}.

To quantitatively characterize the cavity shape, we define two important geometrical parameters: radial elongation $\tau$ and top wideness $b$. We refer cavities with $\tau\leq 1$ (i.e., nearly circular or elongated along the perpendicular direction) as type I cavities, and those with $\tau> 1$ (i.e., elongated along the jet direction) as type II cavities. Both types of X-ray cavities have been observed in the cavity samples of \citet{rafferty06} and \citet{hl12}. The value of $b$ further separates each type of cavities into three subgroups: top-wide ($b>0.5$), center-wide ($b\sim 0.5$), and bottom-wide ($b<0.5$).

Using a suite of 14 axisymmetric hydrodynamic simulations, we study how the cavity shape is affected by various jet properties and viscosity in the ICM. The X-ray cavities in our simulations are directly created by AGN jets, and we also follow their evolution in the ICM. After the jet is turned off, the produced cavity rises in the ICM and its shape continues to evolve. We thus study the shapes of young cavities and old cavities separately to better distinguish the roles of the different physics parameters.

Our simulations indicate that viscosity does not significantly affect the shapes of young cavities, which can be directly used to probe jet properties. We show that it is feasible to produce both type I and II cavities in hydrodynamic simulations. The value of radial elongation $\tau$ increases with the jet density, velocity, and duration, but decreases with the jet energy density and radius. To disentangle the effects of these jet parameters, we suggest to study the second geometrical parameter -- top wideness. Bottom-wide type-I cavities (e.g., the southern inner cavity in Perseus) are produced by very light internally-subsonic jets, while top-wide type-II cavities are produced by heavier jets with internally supersonic velocities, which may also produce center-wide cavities with $\tau\sim 1$ if the jets are slightly supersonic ($M_{\rm int}\sim 1$-$2$). Bottom-wide type-II cavities can be produced by very light jets with very long durations, and cylindrical cavities (both type I and II) are produced by very light internally-supersonic jets. We confirm the above connection between the shapes of young cavities and jet properties with two additional simulations of the Perseus cluster, which is much more massive than our fiducial Virgo cluster.

Viscosity plays a significant role in the long-term evolution of X-ray cavities in the ICM. In non-viscous runs, both type I and II cavities are subject to RT and KH instabilities and the major part of a cavity tends to evolve into a torus-like structure. The role of viscosity becomes more important as the dynamic viscosity coefficient increases. With $\mu_{\rm visc}=100$ g cm$^{-1}$ s$^{-1}$ (a significant fraction of the Braginskii viscosity $\sim156(\text{ln}\Lambda/37)^{-1}(T/2\text{ keV})^{2.5}$ g cm$^{-1}$ s$^{-1}$), viscosity significantly affects the long-term cavity evolution, effectively suppressing both interface instabilities and the formation of torus-like morphology. We also find that the top-wide type-II cavity in run R4 rises much faster than the bottom-wide type-I cavity in run R1, suggesting that the momenta injected by AGN jets, in addition to buoyancy, also contribute significantly to the late rise of top-wide type-II cavities in the ICM. For bottom-wide type-I cavities, buoyancy may play the dominant role, similar to initially static cavities in \citet{reynolds05}.

Our numerical study has important observational implications and we encourage X-ray observers to study the shapes of X-ray cavities. The shapes of young cavities can be used to probe the properties of AGN jets, while the shapes of old cavities may be a useful probe of the ICM viscosity level. One can look at individual cavities. For example, the outer northwestern X-ray cavity in Perseus (see Figure 3 in \citealt{fabian00}) is significantly flattened (cap-shaped) and elongated along the perpendicular direction. In our simulations, the most similar cavity is the one in the viscous run R1-visc2 at $t=100$ Myr (see Figure \ref{plot7}), suggesting that the cavity was created by a very light internally-subsonic jet, and viscosity in the ICM of Perseus is significant. While further studies are necessary to make conclusive statements, the suggested significant viscosity level is consistent with the constraint derived from coherent cold filaments in the wake of this cavity ($\mu_{\rm visc}/\rho>4\times 10^{27}$ cm$^{2}$ s$^{-1}$ or $\mu_{\rm visc}>80$ g cm$^{-1}$ s$^{-1}$ taking $0.01$ cm$^{-3}$ for the electron number density; \citealt{fabian03}). A substantial viscosity level in the ICM is also consistent with the stability of cold fronts observed in some galaxy clusters (e.g., \citealt{zuhone14}). 

Furthermore, it may be even more interesting to statistically study the cavity shape in a large sample of X-ray cavities, investigating if there is a bimodality in the distributions of $\tau$ and $b$, and how the values of $\tau$ and $b$ vary with other cavity properties (e.g., the cavity distance from the cluster center, the mechanical AGN power, etc).

\section*{Acknowledgments}

F.G. acknowledges generous support by the Zwicky Prize Fellowship and the observational cosmology group (PI: Simon Lilly) at ETH Z\"{u}rich. F.G. thanks William Mathews for helpful discussions and two anonymous referees for very insightful comments, which significantly improved the paper.


\label{lastpage}

\end{document}